\documentclass[journal]{IEEEtran}
\usepackage[T1]{fontenc}
\usepackage{amsmath,amsfonts}
\usepackage{array}
\usepackage{graphicx}
\usepackage{cite}
\usepackage{subcaption}

\begin{document}

\title{Learning based Channel Estimation and Phase Noise Compensation in Doubly-Selective Channels} 
\author{Sandesh Rao Mattu and A. Chockalingam \\
\vspace{-4mm}
\thanks{\textcopyright \ 2022 IEEE. Personal use of this material is permitted. Permission from IEEE must be obtained for all other uses, in any current or future media, including reprinting/republishing this material for advertising or promotional purposes, creating new collective works, for resale or redistribution to servers or lists, or reuse of any copyrighted component of this work in other works.
}}

\maketitle
\newcommand{\hhat}{${\bf \hat{H}}\ $}
\newcommand{\htilde}{${\bf \tilde{H}\ }$}
\newcommand{\breveh}{${\bf \breve{H}\ }$}
\newcommand{\blue}[1]{{\color{blue}#1}}

\begin{abstract}
In this letter, we propose a learning based channel estimation scheme for orthogonal frequency division multiplexing (OFDM) systems in the presence of phase noise in doubly-selective fading channels. Two-dimensional (2D) convolutional neural networks (CNNs) are employed for effective training and tracking of channel variation in both frequency as well as time domain. The proposed network learns and estimates the channel coefficients in the entire time-frequency (TF) grid based on pilots sparsely populated in the TF grid. In order to make the network robust to phase noise (PN) impairment, a novel training scheme where the training data is rotated by random phases before being fed to the network is employed. Further, using the estimated channel coefficients, a simple and effective PN estimation and compensation scheme is devised. Numerical results demonstrate that the proposed network and PN compensation scheme achieve robust OFDM performance in the presence of phase noise.
\end{abstract}

\vspace{-2mm}
\section{Introduction}
\label{sec1}
In mobile communication networks, orthogonal frequency division multiplexing (OFDM) systems have to deal with time-selectivity of the channel in addition to frequency-selectivity. A key issue in OFDM is the problem of channel estimation/equalization in time-varying channels \cite{coleri}. In addition to time-selectivity of the channel, OFDM receivers are known to be sensitive to impairments due to local oscillator phase noise \cite{tomba}. With a conventional single-tap equalizer, the bit error rate (BER) performance of OFDM floors due to inter-carrier interference (ICI) caused by phase noise \cite{smaini}. Recently, deep neural networks are finding use in a wide range of problems in the physical layer design, including transceiver designs \cite{shea}-\cite{mehran}. Fully connected neural networks based channel estimation in OFDM systems have been considered in \cite{ce1},\cite{ce2}. These works do not consider the time-selectivity of the fading channel. Convolutional neural network (CNN) is a type of neural network that has been widely used in the field of image and video processing for applications like de-noising and improving the resolution of an image. The application of CNNs for channel estimation in OFDM systems has been considered in \cite{mehran}. In \cite{mehran}, the TF grid of channel coefficients is modelled as an image. The estimates obtained at the pilot locations are interpolated across time and frequency and filtered through a super resolution network and a de-noising network. This approach does not yield a single trained architecture that performs consistently across all signal-to-noise ratio (SNR) values. Also, these works do not consider the problem of channel estimation in the presence of phase noise (PN). Several methods have been reported for PN estimation and compensation, e.g., \cite{ce5},\cite{ce7}. In \cite{ce7}, the authors use learning based schemes to estimate PN and decode symbols in doubly-selective channels. However, the approach assumes the channel to be static over the subframe duration. Further, the computational complexity and pilot density are high. In this work, we propose a novel, low complexity learning based scheme for channel estimation in time-varying channels (where the channel varies in time within one OFDM subframe depending on the Doppler) and phase noise compensation in OFDM systems. Under such rapid time-varying channel conditions, the proposed scheme achieves better performance at lower complexity and pilot density compared to other state-of-the art techniques. The new contributions in this letter are summarized as follows.

We consider the doubly-selective TF channel grid as a 2D image and solve the problem of channel estimation as an image completion problem using sparse data. The sparse data here is the pilot symbols sparsely populated in the TF grid. We propose a 2D CNN architecture for this, which is a natural fit. A novelty here is that, in order to render the network robust to PN, a training methodology where the training data is rotated by random phases before being fed to the network is employed. We further propose a simple time-domain PN compensation scheme that uses the channel estimates from the estimator network and the knowledge of pilot symbols to estimate the PN at the pilot locations, which are then 2D interpolated across the entire TF grid. Simulation results for Vehicular A (VehA) channel model show that the proposed channel estimator network and PN compensation scheme achieve robust mean square error and bit error performance in the presence of PN.

\vspace{-2mm}
\section{System Model}
\label{sys_mod}
Consider a single-input single-output (SISO) OFDM system with $N_c$ subcarriers. Let $\{X_k\}_{k=0}^{N_c-1}$ be the information symbols multiplexed on the $N_c$ subcarriers of one OFDM symbol in the frequency domain. Let the corresponding time domain sequence after inverse discrete Fourier transform (IDFT) be ${\bf x}_t = \{x_n\}_{n=0}^{N_c-1}$. An $N_{\text{cp}}$-length cyclic prefix (CP) is added, and the cyclic-prefixed time domain sequence is transmitted through a frequency-selective channel with $L$ taps ($N_{\text{cp}} \geq L$). The received signal is affected by PN induced multiplicative distortion. The received time domain sequence at the receiver, after removing the CP, is given by  
\begin{align}
{\bf y}_{t} = {\bf \Phi}_t\left({\bf h}_t \text{\textcircled{$*$}}{\bf x}_t + {\bf n}_t\right),
\label{sys1}
\end{align}
where ${\bf \Phi}_t$ = diag$(e^{j\phi_0}, e^{j\phi_1}, \cdots, e^{j\phi_{N_c-1}}) \in \mathbb{C}^{N_c \times N_c}$ is a diagonal matrix with PN realizations on the diagonal, ${\bf h}_t \in \mathbb{C}^{N_c \times 1}$ is the channel impulse response (padded with $N_c-L$ zeros), and ${\bf n}_t \in \mathbb{C}^{N_c \times 1}$ contains i.i.d. circularly symmetric Gaussian noise samples with variance $\sigma^2$. Let {\boldmath $\psi$}$_t = [e^{j\phi_0} \ e^{j\phi_1} \cdots e^{j\phi_{N_c-1}}]^T$ denote the time domain PN vector. The receiver converts ${\bf y}_{t}$ to a frequency domain vector ${\bf y}_{f} \in \mathbb{C}^{N_c \times 1}$ using discrete Fourier transform (DFT), which can be written as  
\begin{align}
{\bf y}_{f} = {\bf \Psi}_f\text{\textcircled{$*$}}\left({\bf h}_f \odot {\bf x}_f + {\bf n}_f\right),
\label{sys2}
\end{align}
where ${\bf \Psi}_f =[\psi_0 \ \psi_1 \cdots \psi_{N_c-1}]^T \in \mathbb{C}^{N_c \times 1}$ represents the DFT coefficient vector of the time domain PN vector {\boldmath $\psi$}$_t$, and ${\bf h}_f,{\bf x}_f,{\bf n}_f \in \mathbb{C}^{N_c \times 1}$ represent the channel response, transmitted symbol, and noise vector in the frequency domain, respectively. Defining a circulant matrix ${\bf \Theta}_f \in \mathbb{C}^{N_c \times N_c}$ as
\begin{align}
{\bf \Theta}_f = 
\begin{bmatrix}
\psi_0 & \psi_{N_c-1} & \cdots & \psi_1 \\
\psi_1 & \psi_0 & \cdots & \psi_2 \\ 
\vdots & \vdots & \ddots & \vdots \\
\psi_{N_c-1} & \psi_{N_c-2} & \cdots & \psi_0 \\
\end{bmatrix},
\label{sys3}
\end{align}
the circular convolution in \eqref{sys2} is equivalently represented as
\begin{align}
{\bf y}_{f} = {\bf \Theta}_f\left({\bf h}_f \odot {\bf x}_f + {\bf n}_f\right).
\label{sys4}
\end{align}
Using the distributive property of matrix multiplication, \eqref{sys4} can be simplified to
\begin{align}
{\bf y}_{f} & \hspace{0.5mm} = \hspace{0.5mm} {\bf \Theta}_f{\bf h}_f \odot {\bf x}_f + {\bf \Theta}_f{\bf n}_f
\hspace{0.5mm} = \hspace{0.5mm} {\bf \tilde{h}}_f \odot {\bf x}_f + {\bf \tilde{n}}_f,
\label{sys5}
\end{align}
where ${\bf \tilde{h}}_f={\bf \Theta}_f{\bf h}_f$ and ${\bf \tilde{n}}_f={\bf \Theta}_f{\bf n}_f$ are the PN effected channel frequency response and noise vectors, respectively, and $({\bf \tilde{n}}_f)_i \sim \mathcal{CN}(0, \sigma^2)$, $i=0, 1, \cdots N_c-1$.

\begin{figure}
\centering
\includegraphics[width=0.8\linewidth]{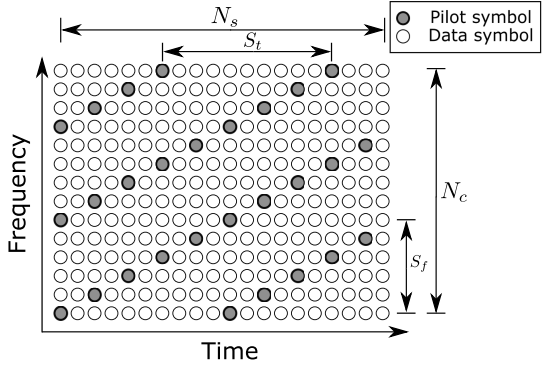}
\caption{Pilot and data symbols placement in the TF grid.}
\label{fig:frame_partition}
\end{figure}

Transmission is divided into subframes. Each subframe consists of $N_s$ OFDM symbols as shown in Fig. \ref{fig:frame_partition}. From \eqref{sys5}, the received signal matrix in the frequency domain corresponding to one subframe can be written as
\begin{align}
{\bf Y}_{f} &= {\bf \tilde{H}}_f \odot {\bf X}_f + {\bf \tilde{N}}_f,
\label{sys6}
\end{align}
where ${\bf X}_f = [{\bf x}_{f}^{(0)} \ {\bf x}_{f}^{(1)}\cdots{\bf x}_{f}^{(N_s-1)}] \in \mathbb{C}^{N_c \times N_s}$ denotes the $N_s$ transmitted OFDM symbols
with the $i$ in ${\bf x}_{f}^{(i)}$ denoting the OFDM symbol index. Likewise, ${\bf \tilde{H}}_f = [{\bf \tilde{h}}_{f}^{(0)} \ {\bf \tilde{h}}_{f}^{(1)} \cdots {\bf \tilde{h}}_{f}^{(N_s-1)}]$ $\in \mathbb{C}^{N_c \times N_s}$, ${\bf \tilde{N}}_f = [{\bf \tilde{n}}_{f}^{(0)} \ {\bf \tilde{n}}_{f}^{(1)} \cdots {\bf \tilde{n}}_{f}^ {(N_s-1)}] \in \mathbb{C}^{N_c \times N_s}$, and ${\bf Y}_f = [{\bf y}_{f}^{(0)} \ {\bf y}_{f}^{(1)} \cdots {\bf y}_{f}^{(N_s-1)}] \in \mathbb{C}^{N_c \times N_s}$ are the channel response, noise, and the received OFDM symbols, respectively.

Towards estimating the coefficients of the channel matrix ${\bf \tilde{H}}_f$ for a subframe, pilot symbols are placed at known locations in the subframe. Figure \ref{fig:frame_partition} shows one such arrangement, called lattice-type pilot arrangement, wherein $N_p$ pilot symbols are placed in the subframe. The pilot symbols are separated in time by $S_t$ time slots and in frequency by $S_f$ subcarriers. Let ${\bf x}_{f, p} \in \mathbb{C}^{N_p \times 1}$ be the vector of transmitted pilot symbols and ${\bf y}_{f, p}$ be the corresponding received vector. Let ${\bf \tilde{h}}_{f, p} \in \mathbb{C}^{N_p \times 1}$ be the vector of channel coefficients seen by the pilot symbols. The vector of least squares (LS) channel estimates at the pilot locations, ${\bf \hat{\tilde{h}}}_{f, p}$, is obtained as
\begin{align}
{\bf \hat{\tilde{h}}}_{f, p} = \underset{{\bf \tilde{h}}_{f, p}}{\text{argmin}} \Vert {\bf y}_{f, p} - {\bf \tilde{h}}_{f, p} \odot {\bf x}_{f, p} \Vert^2,
\label{sys7}
\end{align}
which on solving gives 
\begin{align}
{\bf \hat{\tilde{h}}}_{f, p} = \frac{{\bf y}_{f, p}}{{\bf x}_{f, p}}.
\label{sys8}
\end{align}
Typically, using the knowledge of ${\bf \hat{\tilde{h}}}_{f, p}$ at the pilot locations, interpolation is carried out to obtain the estimates for the entire TF grid, i.e., to obtain an estimate of ${\bf \tilde{H}}_f$. But due to the time-selective nature of the channel and the random nature of rotations introduced by PN, such traditional approaches yield poor estimates/performance. It is therefore necessary to $i)$ learn and track the time variations of the channel, and $ii)$ estimate and compensate the PN in order to achieve robust performance. In the following section, we propose a CNN architecture to solve the first task, and solve the second task using the estimates obtained from the first task.

\section{Proposed channel estimation and PN compensation}
\label{ch_est}
\begin{figure*}
\centering
\includegraphics[width=0.8\linewidth]{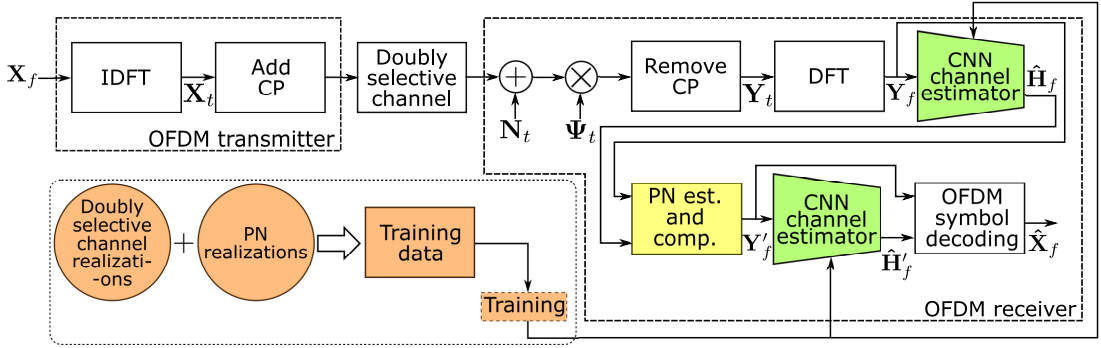}
\caption{Proposed CNN based channel estimator network and PN compensation scheme.}
\label{fig:block_dia}
\end{figure*}

Figure \ref{fig:block_dia} shows the block diagram of the proposed CNN based channel estimator network and PN compensation scheme for OFDM systems. At the transmitter, 
an OFDM subframe comprising of $N_s$ OFDM symbols (with pilot and data symbols as shown in Fig. \ref{fig:frame_partition}), represented by ${\bf X}_f$, is converted to time domain using IDFT operation to obtain ${\bf X}_t \in \mathbb{C}^{N_c \times N_s}$, and prefixed with CP. The subframe is transmitted over a doubly-selective fading channel. The channel matrix for the subframe in the frequency domain is ${\bf H}_f = [{\bf h}_f^{(0)} {\bf h}_f^{(1)} \cdots {\bf h}_f^{(N_s-1)}]$, where ${\bf h}_f^{(i)}$ is the channel response of the $i$th OFDM symbol. The receiver introduces additive noise (${\bf N}_t$) and multiplicative distortion (${\bf \Psi}_t$) due to PN. CP is removed and the resultant matrix ${\bf Y}_t$ is converted to frequency domain using DFT to obtain ${\bf Y}_f$. The matrix ${\bf Y}_f$ comprises of both pilot and data symbols. The pilot symbols are used to obtain an estimate ${\bf \hat{H}}_f = [{\bf \hat{h}}_f^{(0)} {\bf \hat{h}}_f^{(1)} \cdots {\bf \hat{h}}_f^{(N_s-1)}]$  of the channel matrix ${\bf H}_f$ using the proposed channel estimator network.
Following this, ${\bf \hat{H}}_f$ and the received pilot symbols are used to estimate the PN samples and compensate the received subframe ${\bf Y}_f$ to obtain ${\bf Y}'_f$ using the proposed PN compensation algorithm.
A second set of channel estimates, ${\bf \hat{H}}'_f$, is obtained using the channel estimator network with ${\bf Y}'_f$ as the received subframe. Finally, ${\bf \hat{H}}'_{f}$ is used for decoding the data symbols in ${\bf Y}'_f$.

\begin{figure}
\centering
\includegraphics[width=1\linewidth]{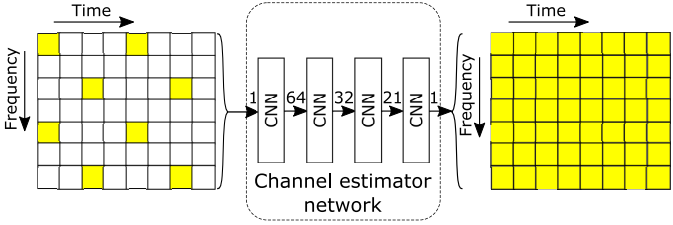}
\caption{Proposed CNN based channel estimator network.}
\label{fig:ch_est}
\end{figure}

\begin{table}
\centering
\begin{tabular}{|c|c|c|c|}
\hline
{\bf Layer} & {\bf Input channels} & {\bf Output channels} & {\bf Kernel size}\\
\hline
1 & 1 & 64 & (16, 4) \\
\hline
2 & 64 & 32 & (16, 4) \\ 
\hline
3 & 32 & 21 & (17, 5) \\ 
\hline
4 & 21 & 1 & (20, 8) \\
\hline
\end{tabular}
\caption{Parameters of the 2D-CNN layers in the channel estimator network.}
\label{tab:ch_est_par}
\end{table}

\subsection{Proposed channel estimator network and training} 
\label{sec3a}
The proposed channel estimator network comprises of four 2D-CNN layers, with the stride of each layer set to one and padding adjusted to make the output of each layer have the same dimension as the input. The input to the network is a sparse TF grid comprising of LS estimates (using \eqref{sys8}) of the channel at the pilot locations and zeros elsewhere. The sparse TF grid is separated into real and imaginary parts and estimation is performed sequentially. Using the sparse information available at the input, the network is trained to `complete' the TF grid, i.e., to provide estimates for the entire TF grid. This is depicted in Fig. \ref{fig:ch_est} wherein, at the input, the squares marked yellow represent the availability of LS estimates in pilot locations and the estimator network provides the estimates for the TF grid at the output, tracking the channel variations in both time and frequency. The other parameters of the channel estimator network are presented in Table \ref{tab:ch_est_par}.

Doubly-selective channel realizations are generated and used for training the channel estimator network. During training, PN induced rotations are introduced at the input of the network. The network is trained using PN samples that are Gaussian distributed \cite{smaini} with zero mean and $\sigma_{{\tiny \mbox{PN}}}^2$ variance. 
Modelling PN samples using Gaussian distribution improves the training robustness as the model is able to learn both with (when PN absolute value is greater than zero) and without (when PN absolute value is close to zero) the effect of PN. This helps the model generalize beyond the $\sigma_{{\tiny \mbox{PN}}}$ values seen while training.
To train the network, the input is set to be a sparse TF-grid, ${\bf H}_p$, where the pilot locations contain channel coefficients of a channel realization with PN and zeros elsewhere. The output of the network is compared against the channel realization, ${\bf H}_{\text{act}}$, using an $L_1$-loss function given by
\begin{align}
L = \overline{\sum_{{\bf H}_{\text{act}}} \vert f({\bf \Theta}_{\text{CE}}, {\bf H}_p) - {\bf H}_{\text{act}} \vert},
\label{l1_loss}
\end{align}
where $f(\cdot)$ represents the channel estimator network, ${\bf \Theta}_{\text{CE}}$ is the set of all trainable parameters in the network, and $\overline{\cdot}$ denotes the mean operation over all the training samples.
The other hyper-parameters used in the training are shown in Table \ref{tab:training_hp}.

\begin{table}
\centering
\begin{tabular}{|l|l|}
\hline
{\bf Hyper-parameter} & {\bf Value} \\
\hline
Epochs & 10000 \\
\hline
Optimizer & Adam \\
\hline
Learning rate & 0.001, divide by 2 every 2000 epochs \\ 
\hline
Batch size & 1000 \\
\hline
Mini-batch size & 64 \\
\hline
\end{tabular}
\caption{Hyper-parameters used for training the channel estimator network.}
\label{tab:training_hp}
\end{table}

\subsection{Proposed PN compensation algorithm}
\label{subsec3b}
The proposed PN compensation algorithm begins by estimating the PN samples in the TF grid, following which the received subframe is compensated. Using the estimate ${\bf \hat{H}}_f$ obtained from the channel estimator network, the $i$th OFDM symbol (consisting of pilot and data symbols) in the received subframe can be approximated as (using \eqref{sys2})
\begin{align}
{\bf y}_{f}^{(i)} = {\bf \Psi}_f^{(i)}\text{\textcircled{$*$}}\left({\bf \hat{h}}_f^{(i)} \odot {\bf x}_f^{(i)} + {\bf n}_f^{(i)} \right).
\label{comp1}
\end{align}
Representing \eqref{comp1} in the time domain yields (the superscript $i$ is dropped for brevity)
\begin{align}
{\bf y}_{t} = \boldsymbol{\psi}_t\left({\bf \hat{h}}_t \text{\textcircled{$*$}}{\bf x}_t + {\bf n}_t\right).
\label{comp2}
\end{align}
Defining a circulant matrix ${\bf \hat{H}}_t$ as
\begin{align}
{\bf \hat{H}}_t = 
\begin{bmatrix}
\hat{h}_t^{(0)} & \hat{h}_t^{(N_c-1)} & \cdots & \hat{h}_t^{(1)} \\
\hat{h}_t^{(1)} & \hat{h}_t^{(0)} & \cdots & \hat{h}_t^{(2)} \\
\vdots & \vdots & \ddots & \vdots \\
\hat{h}_t^{(N_c-1)} & \hat{h}_t^{(N_c-2)} & \cdots & \hat{h}_t^{(0)} \\
\end{bmatrix},
\label{comp3}
\end{align}
\eqref{comp2} can be equivalently written as
\begin{align}
{\bf y}_{t} 
= \boldsymbol{\psi}_t\left({\bf \hat{H}}_t{\bf x}_t + {\bf n}_t\right) = \boldsymbol{\psi}_t\left({\bf \hat{H}}_t{\bf F}^{H}{\bf x}_f + {\bf n}_t\right),
\label{comp4}
\end{align}
where ${\bf F}$ represents the $N_c$-point DFT matrix and ${\bf x}_f$ is the frequency domain vector corresponding to ${\bf x}_t$. Let ${\mathcal J}$ denote the set of subcarrier indices at which pilot symbols are present in ${\bf x}_f$. For indices $j \in {\mathcal J}$, \eqref{comp4} can be written as
\begin{align}
{\bf y}_{t}^{(j \in {\mathcal J})} = \boldsymbol{\psi}_t^{(j \in {\mathcal J})}\left(\left({\bf \hat{H}}_t{\bf F}^{H}{\bf x}_f\right)^{(j \in {\mathcal J})} + {\bf n}_t^{(j \in {\mathcal J})}\right).
\label{comp5}
\end{align}
To obtain an estimate of PN samples at locations indexed by ${\mathcal J}$, the following objective function is minimized:
\begin{align}
\boldsymbol{\hat{\psi}}_t^{(j \in {\mathcal J})} = \underset{\boldsymbol{\psi}_t^{(j \in {\mathcal J})}}{\text{argmin}} \left\Vert {\bf y}_{t}^{(j \in {\mathcal J})} - \boldsymbol{\psi}_t^{(j \in {\mathcal J})}\left({\bf \hat{H}}_t{\bf F}^{H}{\bf x}_f\right)^{(j \in {\mathcal J})} \right\Vert^2.
\label{comp6}
\end{align}
Equation \eqref{comp6} is used on all OFDM symbols containing pilots in the subframe. The PN estimates at all the pilot locations are interpolated across the entire TF grid using an MMSE interpolator to obtain $\boldsymbol{\hat{\Psi}}_t$. For the $i$th received symbol ${\bf y}_t^{(i)}$, a compensated symbol ${\bf y}_t'^{(i)} = \boldsymbol{\hat{\Psi}}^{*(i)}_t{\bf y}_t^{(i)}$ is obtained, where $(\cdot)^*$ indicates the conjugation operation. The compensated time domain symbols are converted to frequency domain to obtain the compensated subframe ${\bf Y}_f'$. A final set of channel estimates, ${\bf \hat{H}}_f'$, is obtained from the pilot locations in ${\bf Y}_f'$ using the channel estimator network again. This is done because ${\bf \hat{H}}_f'$ has higher accuracy when compared to ${\bf \hat{H}}_f$ obtained from ${\bf Y}_f$ having PN induced rotations. ${\bf \hat{H}}'_f$ is used for decoding data symbols in ${\bf Y}'_f$.
The operations outlined in \eqref{comp1} through \eqref{comp6} are carried out using the channel estimates ${\bf \hat{H}}_f$ obtained from the channel estimator network, to estimate the PN values.
The compensated subframe ${\bf Y}_f'$ is obtained using the estimated PN values, and is employed to obtain channel estimates ${\bf \hat{H}}'_f$ from the channel estimator network.

\section{Results and Discussions}
\label{results}
In this section, we present the mean square error (MSE) and BER performance of the proposed channel estimator network and the PN compensation algorithm. For all the simulations presented below, for each subframe, $N_s=14$ and $N_c=72$ as per LTE standards \cite{lte}, $N_p=48, S_f = 6,$ and $S_t=7$. 
50000 realizations of Vehicular A (VehA) channel model defined by ITU-R \cite{itu} with six taps, carrier frequency of $f_c=2.1$ GHz, bandwidth of 1.6 MHz, and user equipment (UE) speed of 50 km/h (corresponding Doppler frequency of $f_D=97$ Hz) are generated. 35000 realizations are used for training the proposed channel estimator network (training data), 5000 realizations are used for validating the training (validation data) and the remaining 10000 realizations are used for testing (test data).
Each tap in the VehA model is Rayleigh distributed with the time selectivity based on Jakes model\cite{jakes}.

For all the results presented below, a single trained channel estimator network is obtained using VehA channel realizations with PN $\sim \mathcal{N}\left(0, \sigma_{\tiny \mbox{PN}}^2\right), \sigma_{\tiny \mbox{PN}} = 1.58^\circ$, and $f_D=97$ Hz as training data. We use PyTorch machine learning library for the implementation, training, and testing of the channel estimator network. We use Nvidia RTX 3090 GPU platform to carry out all the simulations. PN samples are generated from its power spectral density given by \cite{smaini}
\begin{align}
L(f_m) = \frac{B_{\text{PLL}}^2L_0}{B_{\text{PLL}}^2 + f_m^2}\left(1 + \frac{f_{\text{corner}}}{f_m}\right) + L_{\text{floor}},
\label{psd}
\end{align}
where $B_{\text{PLL}}$ is the -3 dB bandwidth of the phase locked loop (PLL), $L_0$ is the in-band phase noise level in rad$^2$/Hz (dBc/Hz), $f_m$ is the frequency offset from the carrier frequency, $f_{\text{corner}}$ is the flicker corner frequency, and $L_{\text{floor}}$ is the noise floor. For performance evaluation, we choose three sets of values for the parameters in \eqref{psd}, that correspond to $\sigma_{\tiny \mbox{PN}}=2.78^\circ, 5.46^\circ$, and $10.85^\circ$, respectively, as shown in Table \ref{tab:phase_noise}.

\begin{table}
\centering
\begin{tabular}{|c|c|c|c|c|c|}
\hline
{\bf Set} & $B_{\text{PLL}}$ & $L_0$  & $f_{\text{corner}}$ & $L_{\text{floor}}$  &$\sigma_{\tiny \mbox{PN}}$ \\
 & (Hz) & (dBc/Hz) & (Hz) & (dBc/Hz) & (degree)\\
\hline
1 & 10$^7$ & -95 & 10$^3$ & -150 & 2.78$^\circ$\\
\hline
2 & $4\times 10^7$ & -95 & 10$^3$ & -150 & 5.46$^\circ$ \\
\hline
3 & $4\times 10^7$ & -89 & 10$^3$ & -150 & 10.85$^\circ$\\
\hline
\end{tabular}
\caption{Phase noise PSD parameters.} 
\label{tab:phase_noise}
\end{table}

\begin{figure}
\centering
\includegraphics[width=9.25cm, height=7.00cm]{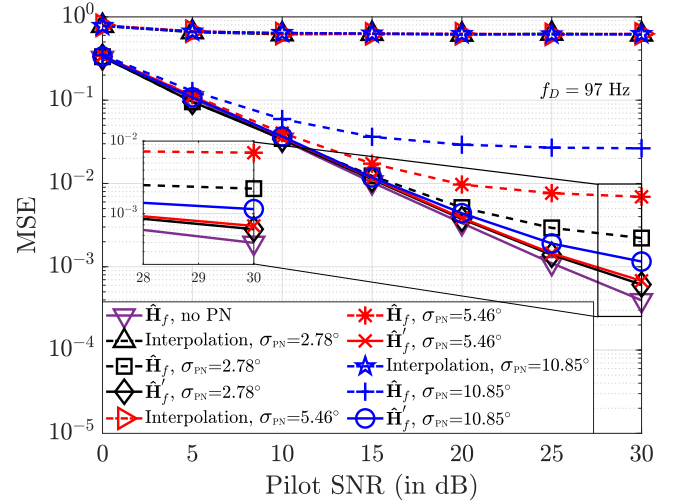}
\caption{MSE performance of the proposed channel estimator network without and with the proposed PN compensation for different values of $\sigma_{\tiny \mbox{PN}}$.}
\label{fig:mse_pn}
\end{figure}

Figure \ref{fig:mse_pn} shows the MSE performance of the proposed channel estimator network as a function of pilot SNR at a Doppler frequency of $f_D=97$ Hz for the three values of $\sigma_{\tiny \mbox{PN}}$ considered. The MSE performance of the first set of channel estimates (${\bf \hat{H}}_f$) and the second set of channel estimates refined by the PN estimation and compensation algorithm (${\bf \hat{H}}_f'$) are plotted. The performance with no PN is also plotted for comparison. In addition, the MSE performance achieved by a 2D spline interpolation scheme is also shown. It is seen that the MSE performance of the interpolation scheme is poor due to the presence of PN and time selectivity of the channel. On the other hand, the proposed network is able to learn and estimate the channel much better. For example, the MSE of the first estimate ${\bf \hat{H}}_f$ itself is much better while the MSE of the second estimate ${\bf \hat{H}}_f'$ is close to the MSE with no PN. This demonstrates the generalization ability of the proposed channel estimator network, wherein the same trained network is able to perform well under different PN levels.

\begin{figure}
    \centering
    \includegraphics[width=9.25cm, height=7.00cm]{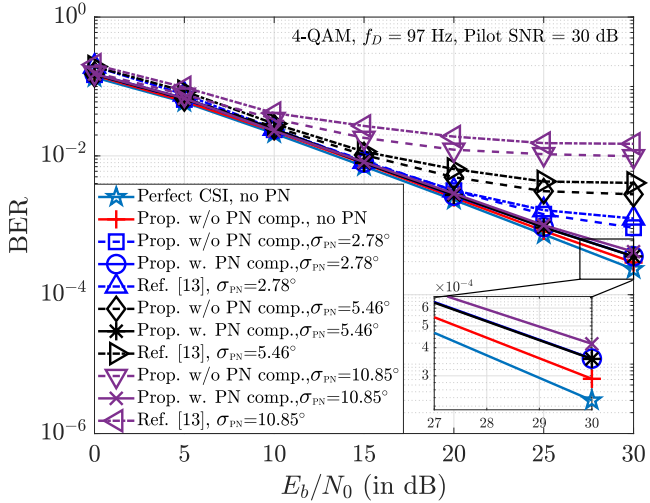}
    \caption{BER performance comparison between the proposed channel estimator network without and with the proposed PN compensation with PN compensation scheme in \cite{sec4a}.}
    \label{fig:comp_trad_meth}
    \vspace{-2mm}
\end{figure}

Next, Fig. \ref{fig:comp_trad_meth} shows the BER performance of the OFDM system as a function of $E_b/N_0$ for 4-QAM, $f_D=97$ Hz, and 30 dB pilot SNR. 
The BER performance is evaluated for the three considered values of $\sigma_{\tiny \mbox{PN}}$ using the channel estimates ${\bf \hat{H}}_f$ (without PN compensation) and ${\bf \hat{H}}_f'$ (with PN compensation). 
For comparison purposes, the BER performance with perfect channel state information (CSI) and no PN is also plotted.
The following observations can be made from the figure. 
Without PN, the BER performance with the proposed channel estimator network is very close to the that with perfect CSI. In the presence of PN, using the first channel estimate ${\bf \hat{H}}_f$, the BER performance floors whereas using the refined estimate ${\bf \hat{H}}_f'$, the performance gets close to that with perfect CSI. For example, while the BER floors at $10^{-2}$ for $\sigma_{\tiny \mbox{PN}}=10.85^\circ$ when ${\bf \hat{H}}_f$ is used, the BER improves to about $4\times 10^{-4}$ at $E_b/N_0=30$ dB for the same $\sigma_{\tiny \mbox{PN}}$ when ${\bf \hat{H}}_f'$ is used. Note that the BER with perfect CSI and no PN for the same $E_b/N_0$ is $2.5\times 10^{-4}$. 
Figure \ref{fig:comp_trad_meth} also presents a comparison of the performance of the proposed scheme with that of the PN compensation scheme (ref. scheme) proposed in \cite{sec4a}.
It is observed that, for all $\sigma_{\tiny \mbox{PN}}$ values, the BER performance of the ref. scheme is comparable with that using the channel estimates $\mathbf{\hat{H}}_f$, while the performance with channel estimates $\mathbf{\hat{H}}_f'$ in the proposed scheme is superior. This can be attributed to the sub-optimality of the iterative scheme followed in \cite{sec4a}. 
Further, the transmitted subframe in \cite{sec4a} consists of an initial block type pilot and comb type pilot thereafter (about 30\% of the symbols are pilots,
while the proposed approach uses lattice-type pilots (only 5\% of symbols are pilots) and has better bandwidth efficiency.

\begin{figure}
\begin{subfigure}{0.5\linewidth}
\centering
\includegraphics[width=4.7cm, height=4.7cm]{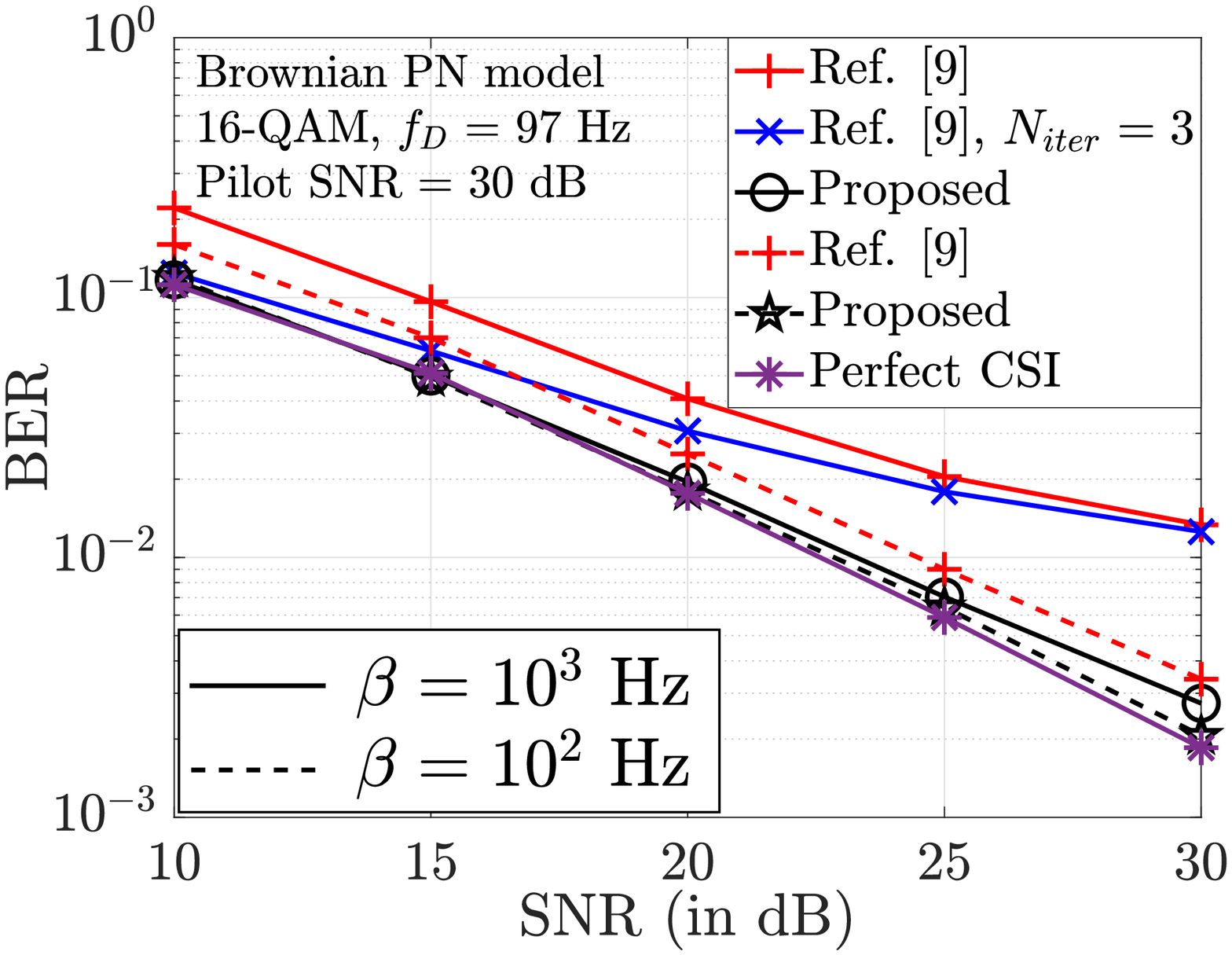}
\caption{BER vs SNR}
\label{fig:comp_state_art}
\end{subfigure}
\begin{subfigure}{0.45\linewidth}
\centering
\includegraphics[width=4.7cm, height=4.7cm]{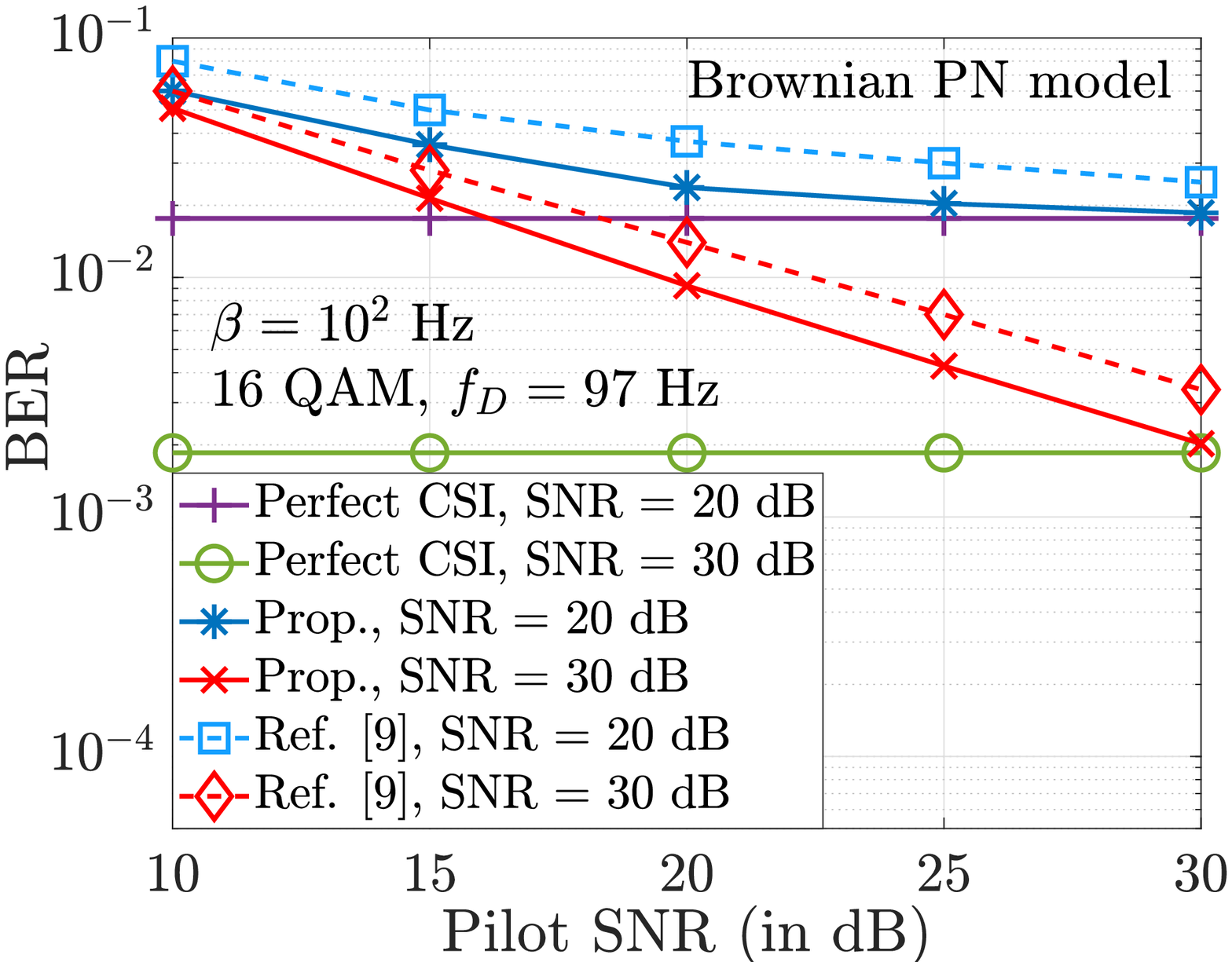}
\caption{BER vs pilot SNR}
\label{fig:ber_vs_pil_snr}
\end{subfigure}
\caption{BER performance comparison between the proposed scheme and the PN compensation scheme in \cite{ce7}.}
\label{fig:perf_state_art}
\vspace{-2mm}
\end{figure}

\subsection{Comparison with NN-based PN compensation in \cite{ce7}
\label{subsec4b}}
Figure \ref{fig:comp_state_art} shows the performance comparison between the proposed channel estimation and PN compensation scheme and the scheme in \cite{ce7} (ref. scheme) for 16-QAM, $f_D=97$ Hz, and 30 dB pilot SNR. The PN is modelled as a Brownian motion process (as in \cite{ce7}) with PN bandwidth parameter $\beta$. The BER performance of the ref. scheme for $\beta = 10^3$ Hz is observed to floor. Further, increasing the number of iterations ($N_{iter}$) to get the initial estimates in the ref. scheme improves its performance in the low and mid SNR regimes. However, the proposed scheme is able to perform better than the ref. scheme throughout the considered SNR range. This can be attributed to the absence of pre-processing in the proposed scheme, while in the ref. scheme the networks are trained using estimates obtained from the first iteration of a non-linear least square estimation algorithm. Next, for $\beta = 10^2$ Hz, the performance of both the schemes are observed to be close, and close to the perfect CSI performance. We note that the ref. scheme considers the pilot arrangement as in \cite{sec4a}, which has lower bandwidth efficiency than the proposed scheme. Also, the ref. scheme is computationally expensive. For example, for each OFDM symbol with $N_c=160$, the ref. scheme involves 4 NNs with about $10^8$ floating point operations (FLOPs), while the proposed scheme's NN requires only about $3\times10^7$ FLOPs.

BER performance as a function of pilot SNR for 20, 30 dB data SNRs are plotted in Fig. \ref{fig:ber_vs_pil_snr}.
As expected, BERs are high at low pilot SNRs due to increased MSE. The gap from respective perfect CSI performance gets more when data SNR is high. This is because when data SNR is high (e.g., at 30 dB), MSE dominates the effect of thermal noise, and vice versa when data SNR is low (e.g., at 20 dB). It is also seen that the proposed scheme performs better than the ref. scheme in \cite{ce7} across all pilot SNRs considered.

\vspace{-2mm}
\section{Conclusions}
\label{conclusions}
We proposed a 2D CNN based learning network to estimate the doubly-selective channel coefficients of a TF grid in an OFDM system by treating the problem as an image completion problem using sparsely available data (i.e., pilot symbols). Numerical results showed that a single trained channel estimator network along with a PN compensation scheme performed well under different PN levels and Doppler frequencies, outperforming other recent schemes.
Compensation of PN considering the effects of interference is suggested as an area for future research.

\end{document}